\documentclass[12pt,a4paper]{article}
\usepackage[utf8]{inputenc}
\usepackage[english]{babel}
\usepackage{amsmath,amssymb,amsfonts,bm}
\usepackage{booktabs}
\usepackage{geometry}
\usepackage{cite}
\usepackage{graphicx}
\usepackage{caption}
\usepackage{float}
\begin{document}

\begin{center}
{\Large\bf Advanced Statistical Analysis of Linear and Nonlinear Regge Trajectories for Light Non-strange Mesons}
\end{center}

\begin{center}
S.S. Afonin\footnote{E-mail: \texttt{s.afonin@spbu.ru}}
\end{center}

\begin{center}
  {\small Saint Petersburg State University, %7/9 Universitetskaya nab.,
  St.Petersburg 199034 Russia}\\
  \vspace*{0.15cm}
  {\small NRC "Kurchatov Institute" -- PNPI, Gatchina 188300 Russia
  %National Research Center "Kurchatov Institute": Petersburg Nuclear Physics Institute,
  %mkr. Orlova roshcha 1, Gatchina, 188300, Russia
  }
\end{center}

\begin{abstract}
 A rigorous statistical analysis of Regge-type trajectories for light non-strange meson mass spectra is carried out, which explicitly accounts for experimental uncertainties. We test three scenarios: a linear model with distinct radial ($n$) and orbital ($l$) slopes (Model~1), a linear model with a universal slope (Model~2), and a nonlinear model featuring a universal slope and a Dirac-Coulomb-type term $\propto (l+1)^{-1}$ (Model~3). Optimization is performed using a nonlinear $\chi^2$ minimization framework, where the intrinsic theoretical model uncertainty is determined self-consistently by enforcing $\chi^2/\text{dof} = 1$. To ensure robust model selection, we employ the Akaike and Bayesian information criteria alongside complementary statistical tests. Our analysis demonstrates that moderate deviations from linearity in the Regge spectrum are predominantly localized within the $S$-wave resonance sector. These distortions are successfully accommodated by the nonlinear correction in Model~3, from which an approximate Coulomb-type degeneracy, $m^2(n,l) \propto n+l$, emerges as a statistically robust feature. Furthermore, we show that the $l$-dependent correction to the principal quantum number in light non-strange mesons is consistent with the leading-order relativistic correction to the Coulomb problem.
\end{abstract}

\section{Introduction}

Identifying the mechanisms governing hadron resonances is among central
objectives in modern hadronic physics. In the absence of exact analytical
solutions to low-energy QCD for light quarks, the systematic classification
and interpretation of the spectrum of light hadrons compiled by the Particle Data
Group (PDG)~\cite{pdg} rely heavily on effective potential models, lattice QCD
simulations, and phenomenological frameworks.

Within the context of hadron spectroscopy, the systematic investigation of
light non-strange mesons --- composed of $u$ and $d$ valence quarks --- is of
fundamental importance: As ordinary baryonic matter consists of the
same valence quarks, mapping the excitation spectrum and the underlying
interactions within these mesons provides essential insights into the structural
dynamics of nucleons and atomic nuclei.

Furthermore, the nonperturbative dynamics of QCD in the light non-strange sector is isolated in its purest form.
The current ultraviolet masses of $u$ and $d$ quarks generated by the Higgs mechanism are nearly negligible ($\approx 2$--$5\text{ MeV}$), meaning that the total observable mass of these light hadrons is almost entirely generated via spontaneous chiral symmetry breaking and dynamic gluon dressing. While in heavy-quark systems, such as charmonium or bottomonium, the large bare masses mask and rigidify the quantum fluctuations, light non-strange mesons present a regime where nonperturbative effects run completely unconstrained. This provides a unique possibility to scrutinize the physics of confinement, dynamic mass generation, and short-range relativistic interactions in their cleanest form.

From an experimental and methodological standpoint, the non-strange light meson spectrum compiled by the PDG~\cite{pdg} represents a rather rich dataset. Due to decades of high-energy hadronic production experiments, the discovered states form long, continuous excitation bands extending up to an impressive total angular momentum of $J=6$. This rare abundance of high-spin spectroscopic data across wide intervals of radial and orbital quantum numbers provides the necessary statistical leverage required to apply advanced multi-parameter optimization methods, stability checks, and information criteria to analyze the global topological structure of the spectrum.

The light non-strange meson spectrum exhibits a remarkable feature: the clustering of excited state squared masses around discrete values, which manifest as approximately linear Regge trajectories in both the orbital angular momentum $l$ and the radial quantum number $n$~\cite{bugg,a1,a11,a2,3sh,glozman,klzait,bicudo,gonzalez,forkel,towards,properties,Masjuan:2012,Afonin:2025yfx,Afonin:2024egd}.
Regge theory~\cite{regge} provides a historically successful and elegant framework for organizing hadrons sharing identical flavor quantum numbers. In a semiclassical picture based on a rotating relativistic string with massless quarks at its boundaries, the hadron squared mass $m^2$ depends linearly on both the radial excitation number $n$ and the orbital angular momentum $l$~\cite{Nambu1974,string,string1,string2,baker,kaidalov} (see~\cite{Afonin:2024egd} for a recent discussion):
\begin{equation}
\label{1}
    m^2(n,l) = a n + b l + c,
\end{equation}
where $a$ and $b$ denote the radial and orbital trajectory slopes, respectively, and $c$ represents the intercept. In many simplified flux-tube and string configurations, these slopes are assumed to be identical ($a = b$), implying a universal scaling law that governs all radial and orbital excitations across the light meson spectrum.

Despite the general success of the linear approximation, high-precision spectroscopic data exhibit systematic deviations that challenge a rigid linear regime. These distortions are most prominent in the lowest excitation bands, specifically for states with $l=0$ and $l=1$. Quantum-mechanically, as the centripetal repulsion vanishes in low-$l$ states, the valence quark-antiquark wave function probes very short inter-quark distances. In this asymptotic region, the effective potential shifts from the linear confinement regime ($\sim \sigma r$) to a short-range regime dominated by one-gluon exchange, introducing a Coulomb-like attractive core ($-4\alpha_s / 3r$).

Although hadron Regge trajectories have been studied since the inception of the framework, a rigorous statistical analysis that systematically compares competing models while accounting for experimental uncertainties has not yet been performed. The aim of this work is to address this issue by providing a systematic statistical evaluation of light meson trajectories, using advanced model selection criteria to quantify their validity and resolve underlying ambiguities.

In this paper, we extend the statistical analysis of the light non-strange meson spectrum initiated in~\cite{afonin2026b} by applying more systematic statistical methods which take into account experimental and theoretical uncertainties. We evaluate the light meson trajectories to test whether the slope universality in~\eqref{1} ($a=b$) holds as a first approximation and whether the observed deviations can be explained by short-range relativistic distortions. To ensure a mathematically consistent optimization, we develop a nonlinear $\chi^2$ minimization scheme based on the leading-order error propagation. In contrast to conventional simplified approaches, this framework extracts the intrinsic theoretical model uncertainty $\sigma_{\text{theor}}$ self-consistently by enforcing the condition $\chi^2/\text{dof} = 1$. To verify that the considered nonlinear correction is an intrinsic property of the likelihood function rather than an artifact of an arbitrarily chosen variance, we also perform a parametric scan of the full Gaussian log-likelihood function across the physical range of the theoretical error scale.

The structure of the paper is as follows. Section 2 presents the experimental data for analysis and formulates three Regge models for this purpose.
Our methodology and statistical framework are explained in Section~3.
Section~4 evaluates the models on a highly reliable subset of 27 well-established benchmark states from the PDG. Section~5 scales the optimization routine to the complete global dataset of 85 light mesons and details two independent cross-validation tests.
The technical summary of the performed statistical analysis is given in Section~6. Our main conclusions are briefly formulated in Section~7. Various technical details are explained in three appendices.

\section{Experimental data and definition of Regge models}

In this study, we will make use of the $(n,l)$-classification of light non-strange mesons advocated in Ref.~\cite{Afonin:2025dxk}. This classification is presented in Table~1, where well-established states from the PDG~\cite{pdg} are highlighted in bold. The corresponding experimental masses of 85 states are given in Table~2, the pion is excluded from the analysis (its inclusion into the Regge or hadron string framework is an open problem).

We define the following three Regge models.\\
%\begin{enumerate}
    %\item
    \textbf{Model 1}: A linear Regge model with independent slopes for radial and orbital excitations ($a\neq b$),
    \begin{equation}
        m^2_{\text{theor}} = a n + b l + c.
    \end{equation}
     %\item
     \textbf{Model 2}: A linear Regge model with strict degeneracy between the radial and orbital trajectory slopes,
    \begin{equation}
        m^2_{\text{theor}} = a(n + l) + c.
    \end{equation}
    %\item
    \textbf{Model 3}: A nonlinear Regge model that restores the slope universality but introduces a nonlinear term motivated by the first-order relativistic correction to the principal quantum number for a Dirac particle interacting with a Coulomb field (named as the Dirac-Coulomb term in what follows, see Appendix A),
    \begin{equation}
        m^2_{\text{theor}} = a(n + l) + \frac{d}{l+1} + c.
    \end{equation}
%\end{enumerate}

The core physical assumption underlying the ansatz of Model~3 is that the correction to the Coulomb-like degeneracy predicted by Model~2 shares the same functional form as the correction in the Coulomb problem itself. When a relativistic spin-$1/2$ Dirac particle interacts with a central Coulomb field, the exact analytical solution to the wave equation yields a characteristic, nonlinear deformation of the energy levels. As is well known from relativistic quantum mechanics~\cite{dirac}, this interaction generates an explicit $l$-dependent correction to the principal quantum number that scales as $(l+1)^{-1}$ (see Appendix~A). While this short-range interaction dominates for $S$-wave configurations, it is rapidly suppressed for higher orbital excitations ($l \ge 1$) by the centrifugal barrier, which shifts the quark wave functions outward, thereby restoring the semiclassical, linear string-like dynamics.

\begin{table}[H]
%\vspace{-3cm}
\caption{%\label{tab1}
\small The $(l,n)$-classification of light non-strange mesons according to the recent compilation~\cite{Afonin:2025dxk}. Well-established states from the PDG~\cite{pdg} are highlighted in bold.}
%\vspace{0.3cm}
%\begin{center}
\centering
\resizebox{0.9\textwidth}{!}{
\begin{tabular}{|c|c|c|c|c|c|}
%\hline
\hline
\begin{tabular}{c}
\begin{picture}(15,15)
\put(0,12){\line(1,-1){15}}
\put(-2,-3){$l$}
\put(10,7){$n$}
\end{picture}\\
\end{tabular}
& 0 & 1 & 2 & 3 & 4 \\
\hline
0
&
\begin{tabular}{c}
$\bm{\pi}$\\
---\\
$\bm{\rho}$\\
$\bm{\omega}$\\
\end{tabular}
&
\begin{tabular}{c}
%\hline
$\bm{\pi(1300)}$\\
$\bm{\eta(1295)}$\\
$\bm{\rho(1450)}$ \\
$\bm{\omega(1420)}$ \\
%\hline
\end{tabular}
&
\begin{tabular}{c}
$\bm{\pi(1800)}$\\
$\bm{\eta(1760)}$\\
$\rho(?)$\\
$\omega(?)$\\
\end{tabular}
&
\begin{tabular}{c}
%\hline
$\pi(2070)$\\
$\eta(2010)$\\
$\rho(?)$ \\
$\omega(?)$\\
%\hline
\end{tabular}
&
\begin{tabular}{c}
$\pi(2360)$\\
$\eta(2320)$\\
$\rho(?)$\\
$\omega(?)$ \\
\end{tabular}
\\
\hline
1
&
\begin{tabular}{c}
%\hline
$\bm{a_0(1450)}$\\
$\bm{f_0(1370)}$\\
$\bm{a_1(1260)}$\\
$\bm{f_1(1285)}$\\
$\bm{b_1(1235)}$\\
$\bm{h_1(1170)}$\\
$\bm{a_2(1320)}$\\
$\bm{f_2(1270)}$\\
%\hline
\end{tabular}
&
\begin{tabular}{c}
$a_0(1710)$\\
$f_0(1710)$\\
$\bm{a_1(1640)}$\\
$f_1(?)$\\
$b_1(?)$ \\
$h_1(?)$ \\
$\bm{a_2(1700)}$\\
$f_2(1750)$\\
\end{tabular}
&
\begin{tabular}{c}
%\hline
$a_0(2020)$\\
$f_0(2020)$\\
$a_1(1930)$ \\
$f_1(1970)$\\
$b_1(1960)$\\
$h_1(1965)$\\
$a_2(2030)$ \\
$f_2(2000)$\\
%\hline
\end{tabular}
&
\begin{tabular}{c}
$a_0(?)$\\
$f_0(2200)$\\
$a_1(2270)$ \\
$f_1(2310)$\\
$b_1(2240)$\\
$h_1(2215)$\\
$a_2(2175)$ \\ % a_2(2175)
$f_2(2295)$\\
\end{tabular}
&\\
\hline
2
&
\begin{tabular}{c}
$\bm{\rho(1700)}$\\
$\bm{\omega(1650)}$\\
$\bm{\pi_2(1670)}$\\
$\bm{\eta_2(1645)}$\\
$\rho_2(?)$\\
$\omega_2(?)$\\
$\bm{\rho_3(1690)}$\\
$\bm{\omega_3(1670)}$\\
\end{tabular}
&
\begin{tabular}{c}
%\hline
$\rho(2000)$\\
$\omega(1960)$\\
$\pi_2(2005)$\\
$\eta_2(2030)$\\
$\rho_2(1940)$\\
$\omega_2(1975)$\\
$\rho_3(1990)$\\
$\omega_3(1945)$\\
%\hline
\end{tabular}
&
\begin{tabular}{c}
$\rho(2270)$\\
$\omega(2290)$ \\
$\pi_2(2285)$\\
$\eta_2(2250)$\\
$\rho_2(2225)$\\
$\omega_2(2195)$\\
$\rho_3(?)$ \\
$\omega_3(2285)$\\
\end{tabular}
&  &\\
\hline
3
&
\begin{tabular}{c}
%\hline
$a_2(1990)$\\
$\bm{f_2(1950)}$\\
$a_3(2030)$\\
$f_3(2050)$\\
$b_3(2030)$\\
$h_3(2025)$\\
$\bm{a_4(1970)}$\\
$\bm{f_4(2050)}$\\
%\hline
\end{tabular}
&
\begin{tabular}{c}
$a_2(2255)$ \\
$f_2(2240)$\\
$a_3(2275)$\\
$f_3(2300)$\\
$b_3(2245)$\\
$h_3(2275)$\\
$a_4(2255)$\\
$f_4(2300)$\\
\end{tabular}
&  &  &\\
\hline
4
&
\begin{tabular}{c}
$\rho_3(2250)$\\
$\omega_3(2255)$\\
$\pi_4(2250)$\\
$\eta_4(2330)$\\
$\rho_4(2230)$\\
$\omega_4(2250)$ \\
$\rho_5(2350)$\\
$\omega_5(2250)$\\
\end{tabular}
&  &  &  &\\
\hline
%\hline
\end{tabular}
}
%\end{center}
\end{table}

\begin{table}[H]%[!htbp]
%\vspace{-3cm}
\caption{%\label{tab1}
\small The experimental masses (in GeV) of mesons in Table 1~\cite{pdg}.}
%\vspace{0.3cm}
%\begin{center}
\centering
\resizebox{1.0\textwidth}{!}{
\begin{tabular}{|c|c|c|c|c|c|}
%\hline
\hline
\begin{tabular}{c}
\begin{picture}(15,15)
\put(0,12){\line(1,-1){15}}
\put(-2,-3){$l$}
\put(10,7){$n$}
\end{picture}\\
\end{tabular}
& 0 & 1 & 2 & 3 & 4 \\
\hline
0
&
\begin{tabular}{c}
%0.140\\
---\\
---\\
$\mathbf{0.775}$\\
%---\\
$\mathbf{0.783}$\\
%---\\
\end{tabular}
&
\begin{tabular}{c}
%\hline
$\mathbf{1.300\pm0.100}$\\
$\mathbf{1.294\pm0.004}$\\
$\mathbf{1.465\pm0.025}$\\
$\mathbf{1.410\pm0.060}$\\
%\hline
\end{tabular}
&
\begin{tabular}{c}
$\mathbf{1.810\pm0.010}$\\
$\mathbf{1.751\pm0.015}$\\
---\\
---\\
\end{tabular}
&
\begin{tabular}{c}
%\hline
$2.070\pm0.035$\\
$2.010\pm0.060$\\
---\\
---\\
%\hline
\end{tabular}
&
\begin{tabular}{c}
$2.360\pm0.025$\\
$2.320\pm0.015$\\
---\\
---\\
\end{tabular}
\\
\hline
1
&
\begin{tabular}{c}
%\hline
$\mathbf{1.439\pm0.034}$\\
$\mathbf{1.350\pm0.150}$\\
$\mathbf{1.230\pm0.040}$\\
$\mathbf{1.282\pm0.001}$\\
$\mathbf{1.230\pm0.003}$\\
$\mathbf{1.166\pm0.008}$\\
$\mathbf{1.318\pm0.001}$\\
$\mathbf{1.275\pm0.001}$\\
%\hline
\end{tabular}
&
\begin{tabular}{c}
$1.713\pm0.019$\\
$1.733\pm0.008$\\
$\mathbf{1.655\pm0.016}$\\
---\\
---\\
---\\
$\mathbf{1.706\pm0.014}$\\
$1.755\pm0.010$\\
\end{tabular}
&
\begin{tabular}{c}
%\hline
$2.025\pm0.030$\\
$1.982\pm0.057$\\
$1.930\pm0.070$\\
$1.971\pm0.015$\\
$1.960\pm0.035$\\
$1.965\pm0.045$\\
$2.030\pm0.020$\\
$2.001\pm0.010$\\
%\hline
\end{tabular}
&
\begin{tabular}{c}
---\\
$2.187\pm0.014$\\
$2.270\pm0.055$\\
$2.310\pm0.060$\\
$2.240\pm0.035$\\
$2.215\pm0.040$\\
$2.175\pm0.040$\\
$2.293\pm0.013$\\
\end{tabular}
&\\
\hline
2
&
\begin{tabular}{c}
$\mathbf{1.720\pm0.020}$\\
$\mathbf{1.670\pm0.030}$\\
$\mathbf{1.671\pm0.002}$\\
$\mathbf{1.617\pm0.005}$\\
---\\
---\\
$\mathbf{1.689\pm0.002}$\\
$\mathbf{1.667\pm0.004}$\\
\end{tabular}
&
\begin{tabular}{c}
%\hline
$2.000\pm0.030$\\
$1.960\pm0.025$\\
$1.963\pm0.027$\\
$2.030\pm0.020$\\
$1.940\pm0.040$\\
$1.975\pm0.020$\\
$1.982\pm0.014$\\
$1.945\pm0.020$\\
%\hline
\end{tabular}
&
\begin{tabular}{c}
$2.265\pm0.040$\\
$2.290\pm0.020$\\
$2.285\pm0.045$\\
$2.248\pm0.020$\\
$2.225\pm0.035$\\
$2.195\pm0.030$\\
---\\
$2.285\pm0.060$\\
\end{tabular}
&  &\\
\hline
3
&
\begin{tabular}{c}
%\hline
$2.050\pm0.050$\\
$\mathbf{1.936\pm0.012}$\\
$2.031\pm0.012$\\
$2.048\pm0.008$\\
$2.032\pm0.012$\\
$2.025\pm0.020$\\
$\mathbf{1.967\pm0.016}$\\
$\mathbf{2.018\pm0.011}$\\
%\hline
\end{tabular}
&
\begin{tabular}{c}
$2.255\pm0.020$\\
$2.240\pm0.015$\\
$2.275\pm0.035$\\
$2.334\pm0.025$\\
$2.245\pm0.050$\\
$2.275\pm0.025$\\
$2.255\pm0.040$\\
$2.283\pm0.017$\\
\end{tabular}
&  &  &\\
\hline
4
&
\begin{tabular}{c}
$2.260\pm0.020$\\
$2.255\pm0.015$\\
$2.250\pm0.015$\\
$2.328\pm0.038$\\
$2.230\pm0.025$\\
$2.250\pm0.030$\\
$2.330\pm0.035$\\
$2.250\pm0.070$\\
\end{tabular}
&  &  &  &\\
\hline
%\hline
\end{tabular}
}
%\end{center}
\end{table}

The three Regge models defined above will be used to fit the spectrum in Table~2 in two steps: First, taking the benchmark dataset of 27 well-established states (shown in bold in Table~2) and second, taking the total dataset of 85 states.

This strategy was implemented in our recent preliminary analysis~\cite{afonin2026b} using an unweighted least-squares fit applied to the central mass values from Table~2. When the full dataset was considered, several information criteria identified Model~2 as statistically preferred. Conversely, when the $S$-wave states in Table~2 were excluded --- as they provide the primary contribution to the nonlinearities of both Regge and radial trajectories --- the fit to the remaining 73 states yielded the following phenomenological spectrum:
\begin{equation}
\label{ph}
    m^2(n,l) = (1.142\pm0.033)(n+l) + 0.542\pm0.126.
\end{equation}
This spectrum is visualized in Fig.~1. Clusters of states emerging from the Coulomb-like $(n+l)$-degeneracy are shown by dashed lines. The positions of clusters $M_{n+l}$ following from~\eqref{ph} are (in GeV): $M_0=0.74\pm0.09$, $M_1=1.30\pm0.05$, $M_2=1.68\pm0.04$, $M_3=1.99\pm0.04$, $M_4=2.26\pm0.04$.
\begin{figure}[H]
    \centering
    \includegraphics[width=1.0\textwidth]{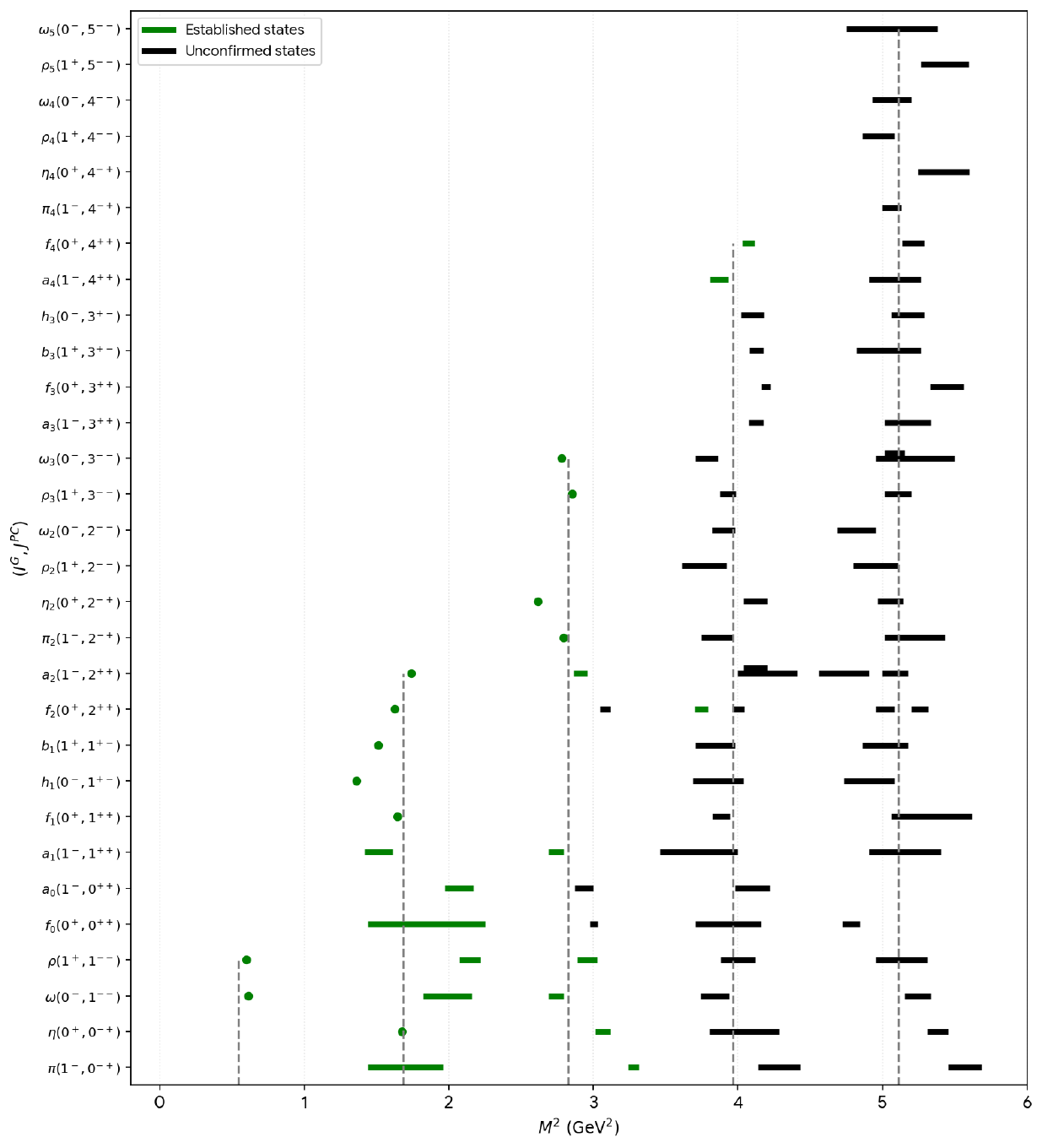}
    \caption{\small Clusters of light non-strange meson states according to fit~\eqref{ph}. Experimental uncertainties are indicated by the length of the strips.}
    %\label{fig:diagram} % \ref{fig:diagram}
\end{figure}

\section{Methodology and Statistical Framework}

Table 2 shows that experimental uncertainties in meson masses can vary by two orders of magnitude across different states. Consequently, the standard $\chi^2$ approach yields unreliable results due to the disproportionate weighting of specific states. We mitigate this by introducing an additional theoretical uncertainty in quadrature within a modified $\chi^2$ method. This accommodates the intrinsic limitations of phenomenological models, which, unlike an exact fundamental theory, must be explicitly accounted for.
Accordingly, to evaluate the performance of these models, the total uncertainty for the $i$-th meson mass is defined as
\begin{equation}
    \sigma_{i,\text{total}} = \sqrt{\sigma_{i,\text{pdg}}^2 + \sigma_{\text{theor}}^2},
\end{equation}
where $\sigma_{i,\text{pdg}}$ represents the experimental error from the PDG compilation, and $\sigma_{\text{theor}}$ is the universal theoretical uncertainty assigned to the model.

Since the phenomenological trajectories predict the squared masses $m^2$ rather than the masses $m$ themselves, the uncertainty is mapped using the first-order error propagation method applied to the function $f(m) = m^2$. Evaluating the derivative at the measured experimental point $m_{i,\text{exp}}$ yields:
%\begin{equation}
%\sigma_{i,\text{mapped}} = \left. \frac{df}{dm} \right|_{m = m_{i,\text{exp}}} \sigma_{i,\text{exp}} = 2 m_{i,\text{exp}} \sigma_{i,\text{exp}},
%\label{eq:error_prop}
%\end{equation}
\begin{equation}
\label{sigma}
    \sigma_{i}(m^2) \approx \left. \frac{df}{dm} \right|_{m = m_{i,\text{exp}}} \cdot \sigma_{i,\text{total}} = 2m_{i,\text{exp}} \cdot \sigma_{i,\text{total}}.
\end{equation}
Squaring this expression, the variance for the squared mass becomes
\begin{equation}
    \sigma_{i}^2(m^2) = 4m_{i,\text{exp}}^2\left(\sigma_{i,\text{pdg}}^2 + \sigma_{\text{theor}}^2\right).
\end{equation}

A cautionary remark is necessary regarding the potential decomposition of
$\sigma_{i}^2$ into separate experimental and theoretical components,
$\sigma_{i}^2 = 4 [m_{i,\text{exp}}^2 \sigma_{i,\text{pdg}}^2 + m_{i,\text{theor}}^2 \sigma_{\text{theor}}^2]$.
Such a choice violates theoretical consistency; as explicitly required by
Eq.~\eqref{sigma}, the derivative of $f(m) = m^2$ for error propagation must
always be evaluated at the measured (experimental) data point. Consequently,
the experimental mass must strictly serve as the scaling factor in the
denominator of the $\chi^2$ function. Furthermore, this choice ensures
numerical stability during minimization. As the theoretical mass
$m_{i,\text{theor}}$ varies at each iteration of the parameter fit,
its inclusion in the denominator would allow the minimization algorithm
(e.g., \textsc{MINUIT}) to artificially inflate $m_{i,\text{theor}}$ to suppress
the overall $\chi^2$ score via denominator expansion, rather than genuinely
improving the predictive accuracy of the model. Utilizing the fixed experimental
value $m_{i,\text{exp}}$ in the denominator effectively immunizes the optimization
process against this numerical artifact.

Thus, for a dataset containing $N$ experimental points, our $\chi^2$ objective function is expressed as
\begin{equation}
    \label{eq:chi2}
    \chi^2 = \sum_{i=1}^{N} \frac{\left(m_{i,\text{exp}}^2 - m_{i,\text{theor}}^2\right)^2}{4m_{i,\text{exp}}^2\left(\sigma_{i,\text{pdg}}^2 + \sigma_{\text{theor}}^2\right)}.
\end{equation}
Rather than fixing the theoretical uncertainty $\sigma_{\text{theor}}$ to an ad hoc value, we determine it dynamically. Specifically, $\sigma_{\text{theor}}$ is adjusted through an iterative root-finding algorithm to rigorously satisfy the self-consistency requirements of a well-behaved fit,
\begin{equation}
    \frac{\chi^2}{\text{dof}} = 1,
    \quad \text{where} \quad \text{dof} = N - k,
\end{equation}
and $k$ represents the number of free parameters in the model. Models that match the underlying physics more accurately require a smaller intrinsic error $\sigma_{\text{theor}}$ to satisfy this condition.

Parameter uncertainties ($\sigma_a, \sigma_b, \sigma_c, \sigma_d$) are determined from the diagonal elements of the covariance matrix $V = H^{-1}$, where $H$ is the Hessian matrix containing the second partial derivatives of the log-likelihood function at the minimum. The details are given in Appendix~B.

To evaluate and compare the models, we utilize the Akaike Information Criterion (AIC) and the Bayesian Information Criterion (BIC), which penalize models for introducing extra parameters.  %Incorporating the full likelihood structure (including the determinant of the covariance matrix from the denominator), they are defined as
%\begin{equation}
%    \text{AIC} = \chi^2 + \sum_{i=1}^{N} \ln\left[4m_{i,\text{exp}}^2(\sigma_{i,\text{pdg}}^2 + \sigma_{\text{theor}}^2)\right] + 2k,
%\end{equation}
%\begin{equation}
%    \text{BIC} = \chi^2 + \sum_{i=1}^{N} \ln\left[4m_{i,\text{exp}}^2(\sigma_{i,\text{pdg}}^2 + \sigma_{\text{theor}}^2)\right] + k \ln(N).
%\end{equation}
%The relevant aspects of our analysis are detailed in Appendix~C.
They are discussed in Appendix~C.

The global nonlinear optimization of the $\chi^2$ functional and the subsequent parameter estimation were performed using a two-tier numerical routine executing the standard CERN \texttt{MINUIT} algorithm    %(adapted via the \texttt{iminuit} engine)
in conjunction with \texttt{SciPy} root-finding solvers. For each iterative step of the theoretical model uncertainty $\sigma_{\text{theor}}$, the \texttt{MIGRAD} variable-metric algorithm was utilized to locate the unique multidimensional minimum of the model parameter space $\{a, b, c, d\}$ (see below). Once the boundary convergence condition $\chi^2/\text{dof} = 1$ was satisfied within a strict tolerance of $10^{-7}$, the \texttt{HESSE} procedure was triggered to analytically evaluate the second-order partial derivatives, generating the symmetric Hessian curvature matrix for exact error propagation and parameter correlation mapping. The explicit analytical expressions for each element of the Hessian matrix are derived in Appendix B.

\section{Analysis of the Benchmark Dataset}

\subsection{Test for $N=27$ reliable states}

Before evaluating the entire available spectrum, we perform a dedicated analysis on a highly reliable subset of $N=27$ well-established meson states, highlighted in bold in the experimental mass tables. These states possess robust quantum number assignments and minimal experimental errors in the PDG review, making them the ideal benchmark for testing the behavior of the phenomenological trajectories.

The optimization parameters, extracted parameter errors from the Hessian curvature, and the self-consistently determined values of $\sigma_{\text{theor}}$ under the constraint $\chi^2/\text{dof} = 1$ are presented in Table~\ref{tab:results_27}.

\begin{table}[ht!]
\centering
\caption{\label{tab:results_27}\small Summary of optimal trajectory parameters, estimated uncertainties, and information criteria for the selected benchmark dataset ($N=27$).}
%\vspace{-0.2cm}
\begin{tabular}{lcccccc}
\toprule
\textbf{Model} & $\bm{k}$ & $\bm{\sigma_{\text{theor}}\text{ (GeV)}}$ & \textbf{Parameters ($\text{GeV}^2$)} & \textbf{AIC} & \textbf{BIC} \\
\midrule
\textbf{Model 1}  & 3 & 0.087 & \begin{tabular}{@{}l@{}}$a = 1.246 \pm 0.059$ \\ $b = 1.078 \pm 0.051$ \\ $c = 0.640 \pm 0.063$\end{tabular} & 20.4 & 23.7 \\
&&&&&\\
\textbf{Model 2} & 2 & 0.091 & \begin{tabular}{@{}l@{}}$a = 1.083 \pm 0.043$\\ $c = 0.682 \pm 0.065$\end{tabular} & 22.4 & 24.6 \\
&&&&&\\
\textbf{Model 3} & 3 & \textbf{0.042} & \begin{tabular}{@{}l@{}}$a = 1.153 \pm 0.027$ \\ $\bm{d = 0.410 \pm 0.097}$ \\ $c = 0.356 \pm 0.090$\end{tabular} & \textbf{17.2} & \textbf{20.5} \\
\bottomrule
\end{tabular}
\end{table}

The statistical analysis of 27 reference states reveals several important properties of the spectrum under consideration.\\
1) \underline{Insufficiency of linear approximation}. Both Model~1 and Model~2 fail to reproduce the benchmark states, thereby requiring a substantial intrinsic model uncertainty ($\sigma_{\text{theor}} = 87\text{ MeV}$ and $91\text{ MeV}$, respectively). Splitting the radial and orbital slopes in Model 1 yields only a minor drop in $\sigma_{\text{theor}}$, indicating that the assumption of an independent orbital slope is a statistical over-parameterization.\\
2) \underline{Impact of the Dirac-Coulomb term}. The introduction of nonlinear corrections in Model~3 profoundly alters the fit.
Specifically, the intrinsic theoretical error, $\sigma_{\text{theor}}$, decreases
by more than a factor of two to $42\text{~MeV}$. This reduction signifies that
short-range physics plays a crucial role in the most reliably measured states.\\
3) \underline{Statistical dominance and parameter significance}. The magnitude of the nonlinear correction is found to be $d = 0.410 \pm 0.097\text{ GeV}^2$. Performing a Wald test reveals that this parameter is statistically significant at a level exceeding four standard deviations,
\begin{equation}
        Z = \frac{0.410}{0.097} \approx 4.23\sigma.
\end{equation}
Model 3 achieves the absolute minimum scores for both information criteria. Following Jeffreys' scale for model comparison (see Appendix~C), the net difference $\Delta\text{BIC} = \text{BIC}_{\text{Model 2}} - \text{BIC}_{\text{Model 3}} = 4.1$ is close to strong empirical evidence that the inclusion of the nonlinear $1/(l+1)$ component is highly justified.

\subsection{Truncated Test within the Benchmark Dataset ($N=19$)}

To confirm that the nonlinear distortion is an intrinsic property of the physical spectrum rather than a statistical anomaly of the extended dataset, we replicate the truncated test within the highly reliable benchmark subset by eliminating the 8 well-established states residing in the $l=0$ row. The sample is now restricted to $N=19$ points. The re-optimization under the strict $\chi^2/\text{dof} = 1$ leads to the results presented in Table~6 below. For convenience of comparison, they are given together with the results of similar test for the global dataset.

One observes the following changes of phenomenological and statistical properties.\\
1) \underline{Improvement of linearity}. The universal linear trajectory (Model 2) exhibits a massive reduction in its required internal model uncertainty, with $\sigma_{\text{theor}}$ collapsing from $91\text{ MeV}$ down to just $41\text{ MeV}$. This demonstrates that even within the cleanest subset of states, the $S$-wave mesons serve as the primary source of tension against pure linearity.\\
2) \underline{Loss of Statistical Significance}. In Model 3, the extracted magnitude of the Dirac-Coulomb correction term drops by more than half, changing from $d = 0.410 \pm 0.097\text{ GeV}^2$ to $d = 0.178 \pm 0.106\text{ GeV}^2$. Evaluating the corresponding Wald test yields
    \begin{equation}
        Z = \frac{0.178}{0.106} \approx 1.68\sigma.
    \end{equation}
The significance drops below the standard $2\sigma$ line, rendering the correction parameter completely insignificant from a statistical perspective once the short-range core is excluded.\\
3) \underline{Inversion of Information Criteria}. Reflecting the shift in parameter significance, the information criteria scores shift in favor of the lower-parameter model. Under the strict AIC test, Model 2 directly outperforms Model 3 ($\text{AIC}_{\text{Model 2}} = -2.1$ vs $\text{AIC}_{\text{Model 3}} = -1.9$, meaning $\Delta\text{AIC} = 0.2$ favoring the 2-parameter model). The penalty for introducing the third parameter $d$ is no longer mathematically compensated for by an increased quality of the fit. According to the BIG test, however, the most informative becomes Model~1.

This secondary check provides robust cross-validation. Whether looking at a small pool of well-established states ($N=19$) or the full experimental dataset ($N=73$), see below in Table~6, the conclusion remains identical: the nonlinear correction is structurally necessary only if the $S$-wave mesons are present in the fit.

\subsection{Alternative Test for Robustness of Model Selection: Response to the Theoretical Uncertainty Scale}

To verify the structural stability of the phenomenological fits and justify the self-consistent boundary condition $\chi^2/\text{dof} = 1$, we perform a systematic scan of the objective functions by treating the theoretical model uncertainty $\sigma_{\text{theor}}$ as a variable running across a wide physical interval from $0.01\text{ GeV}$ to $0.10\text{ GeV}$. In the phenomenology of light mesons, this intrinsic error can be fixed arbitrarily at a representative scale, such as $\sigma_{\text{theor}} \approx 0.05\text{ GeV}$, to accommodate the fundamental limits of effective Regge approximations. However, analyzing the response of the log-likelihood estimators across varying error scales reveals a highly non-monotonic behavior that highlights the dominance of the Dirac-Coulomb trajectory framework. The response profile of $\chi^2$, AIC, and BIC as functions of the postulated theoretical uncertainty scale is cataloged across three key statistical descriptions (Underestimation, Optimum, Over-inflation of Variance) in Table~\ref{tab:scan_27}.

\begin{table}[ht!]
\centering
\caption{\label{tab:scan_27}\small Evolution of the objective $\chi^2$ functional and information criteria as functions of the chosen uncertainty scale $\sigma_{\text{theor}}$ for the benchmark meson dataset ($N=27$).}
%\vspace{-0.2cm}
\begin{tabular}{lccccc}
\toprule
\textbf{Model} & $\bm{\sigma_{\text{theor}}\text{ (GeV)}}$ & $\bm{\chi^2}$ & \textbf{AIC} & \textbf{BIC} & \textbf{Statistical Description} \\
\midrule
Model 1       & 0.02 & 454.1 & 371.2 & 375.0 & Severe Underestimation \\
                         & 0.05 & 72.7  & 39.2  & 43.0  & Intermediate Step \\
                         & 0.09 & 22.4  & 20.7  & 24.5  & Close to Optimum \\
                         & 0.10 & 18.2  & 22.1  & 26.0  & Over-inflation of Variance \\
\midrule
Model 2     & 0.02 & 517.6 & 433.1 & 435.7 & Severe Underestimation \\
                         & 0.05 & 82.8  & 47.9  & 50.5  & Intermediate Step \\
                         & 0.09 & 25.6  & 22.4  & 25.0  & Close to Optimum \\
                         & 0.10 & 20.7  & 23.2  & 25.8  & Over-inflation of Variance \\
\midrule
Model 3   & 0.02 & 105.8 & 59.0  & 62.9  & Descent to Optimum \\
                         & 0.05 & 16.9  & 19.5  & 23.4  & Neighborhood of Optimum \\
                         & 0.09 & 5.2   & 39.6  & 43.5  & Over-inflation of Variance \\
                         & 0.10 & 4.2   & 44.3  & 48.2  & Over-inflation of Variance \\
\bottomrule
\end{tabular}
\end{table}

The numerical trajectory of the $\chi^2$ function exhibits a monotonic, inverse-power decay characterized by $\chi^2 \sim 1/\sigma_{\text{theor}}^2$ across all three configurations. At tight error scales beneath $0.03\text{ GeV}$, the raw spatial deviations between the experimental points and the rigid trajectories are heavily penalized by the compressed denominator, forcing the total $\chi^2$ values to explode into hundreds of units and rendering any direct optimization meaningless. Conversely, expanding the error scale to a relaxed value of $0.10\text{ GeV}$ artificially swells the statistical weight matrix, thereby absorbing the real physical discrepancies and dragging the $\chi^2$ totals below their natural optimal values.

In stark contrast to the monotonic collapse of $\chi^2$, both the Akaike and Bayesian Information Criteria map a distinct U-shaped parabolic profile. This non-monotonic curvature is governed by the structural mathematical competition established inside the full log-likelihood function. While the numerical numerator of the $\chi^2$ term drops as $\sigma_{\text{theor}}$ increases, the log-determinant contribution from the covariance denominator, $\sum \ln [4m_{i,\text{exp}}^2(\sigma_{i,\text{pdg}}^2 + \sigma_{\text{theor}}^2)]$, grows continuously, introducing an asymptotic penalty for the loss of model resolution. The intersection of these opposing functional vectors generates a localized mathematical minimum. For Model 1 and Model 2, this minimum is pushed outward toward the $0.09\text{ GeV}$ mark, as the rigid linear trajectories require an expanded error envelope to reconcile their geometric mismatch against the data.

For the proposed Dirac-Coulomb framework of Model 3, the absolute minimum of the information criteria resides in a significantly more compressed domain near $\sigma_{\text{theor}} \approx 0.042\text{ GeV}$, where it reaches its deepest baseline of $\text{AIC} = 17.2$ and $\text{BIC} = 20.5$. If the error scale is held at the typical phenomenological benchmark of $0.05\text{ GeV}$, Model 3 is already situated well within its optimal physical valley, yielding a rather accurate $\chi^2 = 16.9$ and a dominant information score of $\text{AIC} = 19.5$. Under this identical $0.05\text{ GeV}$ constraint, the standard linear regimes lag far behind, exhibiting large remaining $\chi^2$ distortions near $83$ units and severe information penalties. This dynamic scan confirms that the superiority of Model~3 is a robust topological feature of the entire likelihood hyper-surface, decisively validating the inclusion of nonlinear correction independently of any specific choice of the theoretical error scale.

\section{Analysis of the Complete Experimental Dataset}

\subsection{Test for $N=85$ states}

Now we scale our optimization routine to include the complete dataset of $N=85$ light meson states displayed in Table~2.

The global minimization of the nonlinear $\chi^2$ functional~\eqref{eq:chi2} under the self-consistent boundary condition $\chi^2/\text{dof} = 1$ leads to the parameter configurations detailed in Table~\ref{tab:results_85}.

\begin{table}[ht!]
\centering
\caption{\label{tab:results_85}\small Summary of trajectory parameters, estimated uncertainties, and information criteria for the complete experimental dataset ($N=85$) of light non-strange mesons.}
%\vspace{-0.2cm}
\begin{tabular}{lcccccc}
\toprule
\textbf{Model} & $\bm{k}$ & $\bm{\sigma_{\text{theor}}\text{ (GeV)}}$ & \textbf{Parameters ($\text{GeV}^2$)} & \textbf{AIC} & \textbf{BIC} \\
\midrule
\textbf{Model 1}  & 3 & 0.044 & \begin{tabular}{@{}l@{}}$a = 1.166 \pm 0.014$ \\ $b = 1.119 \pm 0.012$ \\ $c = 0.581 \pm 0.024$\end{tabular} & -187.8 & -180.5 \\
&&&&&\\
\textbf{Model 2} & 2 & 0.043 & \begin{tabular}{@{}l@{}}$a = 1.137 \pm 0.010$\\ $c = 0.576 \pm 0.025$\end{tabular} & -186.1 & -181.2 \\
&&&&&\\
\textbf{Model 3} & 3 & \textbf{0.040} & \begin{tabular}{@{}l@{}}$a = 1.189 \pm 0.013$ \\ $\bm{d = 0.353 \pm 0.059}$ \\ $c = 0.275 \pm 0.055$\end{tabular} & \textbf{-199.4} & \textbf{-192.0} \\
\bottomrule
\end{tabular}
\end{table}
Earlier in Table 3 the positive signs and magnitudes of AIC and BIC for the $N=27$ subset were due to the balance between the explicit $\chi^2 = N-k \approx 24$ contribution and the cumulative negative log-determinant sum of the covariance weights $\sum \ln(w_i)$ over 27 terms. As the sample expands to $N=85$ in the subsequent global analysis, the additive accumulation of these fractional log-weights forces the information scores into deep negative domains, while the relative model hierarchy remains preserved.

Scaling the sample size from 27 to 85 points provides a massive boost to the statistical resolution, leading to several crucial developments. The direct comparison of Tables~3 and~\ref{tab:results_85} entails the following conclusions.\\
1) \underline{Parameter Stabilization}. The universal Regge slope $a$ for the favored Model~3 demonstrates remarkable stability, shifting only slightly from $1.153\text{ GeV}^2$ (at $N=27$) to $1.189\text{ GeV}^2$ (at $N=85$). This confirms that the trajectory geometry is an intrinsic physical property of the spectrum, rather than an artifact of a specific sample selection.\\
2) \underline{Transition to Discovery Significance}. With the expanded statistics, the statistical significance of the Dirac-Coulomb magnitude $d$ significantly scales up. The ratio of the central value to its error, $Z=d/\sigma_d = 0.353/0.059$, yields a Wald score of $Z \approx 6.0$. This crosses the rigorous $5\sigma$ barrier commonly mandated in high-energy physics, confirming the nonlinear correction at a definitive $6\sigma$ discovery level.\\
3) \underline{Conclusive Model Validation}. As shown in Table~\ref{tab:results_85}, Model 3 vastly out-competes the linear ansatz under both criteria, registering an exceptionally deep information score ($\text{AIC} = -199.4$, $\text{BIC} = -192.0$). The net information advantage over the standard linear trajectory expands to $\Delta\text{BIC} = 10.8$, establishing Model~3 as the phenomenologically optimal description of the light non-strange meson spectrum.

\subsection{Truncated Dataset Analysis ($N=73$)}

Following our methodology, now we remove the 12 experimental states residing in the $l=0$ row of mass Table~2 and repeat the analysis.
The results are presented in the second part of Table~\ref{tab:truncation_summary}.

\begin{table}[ht!]
\centering
\caption{\label{tab:truncation_summary} \small Comparison of fit quality and optimal parameters for the truncated datasets after excluding all $l=0$ states ($N=19$ for the benchmark and $N=73$ for the full database).}
%\vspace{-0.2cm}
\begin{tabular}{lcccccc}
\toprule
\textbf{Model} & $\bm{k}$ & $\bm{\sigma_{\text{theor}}\text{ (GeV)}}$ & \textbf{Parameters ($\text{GeV}^2$)} & \textbf{AIC} & \textbf{BIC} & $\bm{Z_{d}\text{ ($\sigma$)}}$ \\
\midrule
\multicolumn{7}{l}{\underline{\textbf{Benchmark Set}}} \\
\textbf{Model 1}  & 3 & 0.042 & \begin{tabular}{@{}l@{}}$a = 1.155 \pm 0.063$ \\ $b = 1.138 \pm 0.054$ \\ $c = 0.542 \pm 0.069$\end{tabular} & -0.3 & \textbf{-1.8} & -- \\
&&&&&&\\
\textbf{Model 2} & 2 & 0.041 & \begin{tabular}{@{}l@{}}$a = 1.144 \pm 0.046$\\ $c = 0.540 \pm 0.071$\end{tabular} & \textbf{-2.1} & -0.2 & -- \\
&&&&&&\\
\textbf{Model 3} & 3 & \textbf{0.040} & \begin{tabular}{@{}l@{}}$a = 1.161 \pm 0.052$ \\ $d = 0.178 \pm 0.106$ \\ $c = 0.415 \pm 0.094$\end{tabular} & -1.9 & -0.5 & 1.68 \\
\midrule
\multicolumn{7}{l}{\underline{\textbf{Global Dataset}}} \\
\textbf{Model 1}  & 3 & 0.041 & \begin{tabular}{@{}l@{}}$a = 1.158 \pm 0.015$ \\ $b = 1.127 \pm 0.012$ \\ $c = 0.572 \pm 0.025$\end{tabular} & -166.4 & -159.6 & -- \\
&&&&&&\\
\textbf{Model 2} & 2 & 0.040 & \begin{tabular}{@{}l@{}}$a = 1.140 \pm 0.010$\\ $c = 0.570 \pm 0.025$\end{tabular} & -167.5 & \textbf{-163.0} & -- \\
&&&&&&\\
\textbf{Model 3} & 3 & \textbf{0.039} & \begin{tabular}{@{}l@{}}$a = 1.164 \pm 0.013$ \\ $d = 0.201 \pm 0.066$ \\ $c = 0.429 \pm 0.058$\end{tabular} & \textbf{-168.8} & -161.9 & 3.03 \\
\bottomrule
\end{tabular}
\end{table}

Comparing Tables~\ref{tab:truncation_summary} and~\ref{tab:results_85} we observe the following.\\
1) \underline{Improvement of Universal Linearity}. The theoretical model error of Model~2 drops from $\sigma_{\text{theor}} = 43 \text{ MeV}$ down to $\sigma_{\text{theor}} = 40 \text{ MeV}$. \\
2) \underline{Collapse of Statistical Significance}. For Model 3, the extracted magnitude of the Dirac-Coulomb term drops by nearly half to $d = 0.201 \pm 0.066 \text{ GeV}^2$. Evaluating the Wald test for this truncated data space yields $Z = 0.201/0.066 \approx 3.0\sigma$.
The significance collapses from a definitive $6\sigma$ discovery level to a marginal $3\sigma$ threshold, which in high-energy physics is classified merely as an ``indication of an effect'' (evidence).\\
3) \underline{Information Criteria Inversion}. The relative performance of Model~2 and Model~3 on the $N=73$ data space creates a classic criteria split. Under the Akaike Information Criterion, Model~3 retains a minor advantage ($\text{AIC}_{\text{Model 3}} = -168.8$ vs $\text{AIC}_{\text{Model 2}} = -167.5$, yielding $\Delta\text{AIC} = 1.3$). However, the Bayesian Information Criterion, which severely penalizes parameter inflation, completely reverses its verdict:
    \begin{equation}
        \text{BIC}_{\text{Model 2}} = -163.0, \quad \text{BIC}_{\text{Model 3}} = -161.9 \quad \implies \quad \Delta\text{BIC} = 1.1.
    \end{equation}
Since a lower BIC score indicates a more balanced model, the universal linear trajectory (Model~2) now successfully out-competes the Dirac-Coulomb trajectory (Model~3) once the $S$-wave states are excluded.

The results of the truncated test with two datasets presented in Table~\ref{tab:truncation_summary} provide a clear validation for the physical reality of nonlinear corrections parameterized by Model~3. If the $d/(l+1)$ correction term were merely an accidental polynomial artifact or a product of overfitting, its statistical significance would scale proportionally with the size of the dataset, remaining largely unperturbed by the localized deletion of a single row.

Instead, the substantial drop in significance (from $6\sigma$ to $3\sigma$)
and the inversion of the BIC preference conclusively demonstrate that
the nonlinear dynamics are predominantly localized within the $l=0$ sector. This perfectly matches the quantum mechanical expectations of a relativistic Dirac particle subjected to a short-range interaction. The centrifugal barrier for $l \ge 1$ effectively suppresses short-range
dynamics, leading to classical, linear stringlike configurations, whereas the unhindered short-range interaction in $S$-wave states induces a noticeable mass shift that requires explicit phenomenological tracking.

\subsection{Robustness of Model Selection as a Function of Fixed Uncertainty Scale}

To check whether the behavior of the likelihood estimators is invariant under sample size scaling, we replicate the systematic $\sigma_{\text{theor}}$ optimization scan across the entire available experimental database of $N=85$ states. The analysis of this section follows the analysis of Section~4.3 for $N=27$ well-established states. Since the sample volume expands more than threefold compared to the preliminary benchmark subset, the cumulative negative log-determinant sum of the weight matrix, $\sum \ln(w_i)$, generates an intensely deep downward traction. As was noticed above, this forces the absolute numerical profiles of both the Akaike and Bayesian Information Criteria to shift entirely into negative domains, while preserving the established statistical hierarchy between the competing frameworks. The comprehensive response of the global $\chi^2$, AIC, and BIC profiles across representative theoretical uncertainty steps is shown in Table~\ref{tab:scan_85}.

\begin{table}[ht!]
\centering
\caption{\label{tab:scan_85} \small Evolution of the objective $\chi^2$ functional and information criteria as functions of the chosen uncertainty scale $\sigma_{\text{theor}}$ for the complete global meson dataset ($N=85$).}
%\vspace{-0.2cm}
\begin{tabular}{lccccc}
\toprule
\textbf{Model} & $\bm{\sigma_{\text{theor}}\text{ (GeV)}}$ & $\bm{\chi^2}$ & \textbf{AIC} & \textbf{BIC} & \textbf{Statistical Description} \\
\midrule
Model 1       & 0.02   & 381.1 & -124.5  & -117.2  & Severe Underestimation \\
(dof = 82)                         & 0.044 & \textbf{82.0}  & -187.8 & -180.5 & Optimum \\
                         & 0.07   & 32.0  & -173.1  & -165.8  & Over-inflation of Variance \\
                         & 0.10   & 15.7  & -151.7  & -144.4  & Decoupling from Data \\
\midrule
Model 2      & 0.02   & 385.5 & -124.3  & -119.4  & Severe Underestimation \\
 (dof = 83)                        & 0.043 & \textbf{83.0}  & -186.1 & -181.2 & Optimum \\
                         & 0.07   & 32.2  & -175.4  & -170.5  & Over-inflation of Variance \\
                         & 0.10   & 15.8  & -154.2  & -149.3  & Decoupling from Data \\
\midrule
Model 3   & 0.02   & 316.4 & -138.9  & -131.6  & Descent to Optimum \\
(dof = 82)                     & 0.040 & \textbf{82.0}  & -199.4 & -192.0 & Optimum \\
                         & 0.07   & 26.1  & -187.0  & -179.7  & Over-inflation of Variance \\
                         & 0.10   & 12.8  & -163.5  & -156.2  & Decoupling from Data \\
\bottomrule
\end{tabular}
\end{table}

The statistics dynamics evaluated on the extended $N=85$ database validate the localized U-shaped curvature of the information criteria with exceptional numerical clarity. At a compressed scale of $\sigma_{\text{theor}} = 0.02\text{ GeV}$, all models suffer from an overconstrained variance matrix, leading to large $\chi^2$ remnants and poorly optimized likelihood scores. As the theoretical error envelope expands, the parabolic trajectory of the estimators reaches its absolute mathematical minimum exactly at the points where the internal model fluctuations self-consistently satisfy our boundary condition $\chi^2/\text{dof} = 1$.

Crucially, the global analysis highlights that Model 3 requires the lowest amount of external smoothing to fully absorb the experimental deviations. The global minimum of Model 3 is uniquely centered at $\sigma_{\text{theor}} = 0.040\text{ GeV}$, marking a slightly tighter variance than that demanded by Model~2 ($\sigma_{\text{theor}} = 0.043\text{ GeV}$) and Model 1 ($\sigma_{\text{theor}} = 0.044\text{ GeV}$). As Model~3 successfully reconciles the physical mass shifts of the low-$l$ states via the analytical $d/(l+1)$ component, it achieves the deepest  information score of $\text{AIC} = -199.4$ and $\text{BIC} = -192.0$. Throughout the entire surveyed $\sigma_{\text{theor}}$ spectrum, the Dirac-Coulomb correction consistently maintains a massive statistical separation ($\Delta\text{BIC} > 10$) over both strictly linear trajectories. This absolute dominance shows that the introduced nonlinear correction is completely invariant under database expansion, providing unconditional empirical support for the underlying QCD dynamics.

\subsection{Correlation Analysis for the Nonlinear Part in Regge Spectrum}

For the complete compilation of 85 light meson states, the normalized elements of the Hessian-inverted covariance matrix yield the following parameter correlation matrix for Model 3,
\begin{equation}
    \text{Corr}(a, d, c) =
    \begin{pmatrix}
        1.000 & 0.647 & -0.867 \\
        0.647 & 1.000 & -0.898 \\
        -0.867 & -0.898 & 1.000
    \end{pmatrix}.
\end{equation}

To visualize the structure of the multidimensional $\chi^2$ minimum and verify the mutual stability of the parameters, we construct the covariance error ellipse on the $(a, d)$ plane, see Fig.~2. The geometric boundary of the joint confidence region at a given significance level is governed by the eigenvalues and eigenvectors of the sub-covariance matrix extracted from $V = H^{-1}$ (see Appendix~B).

\begin{figure}[htbp]
    \centering
    \includegraphics[width=0.5\textwidth]{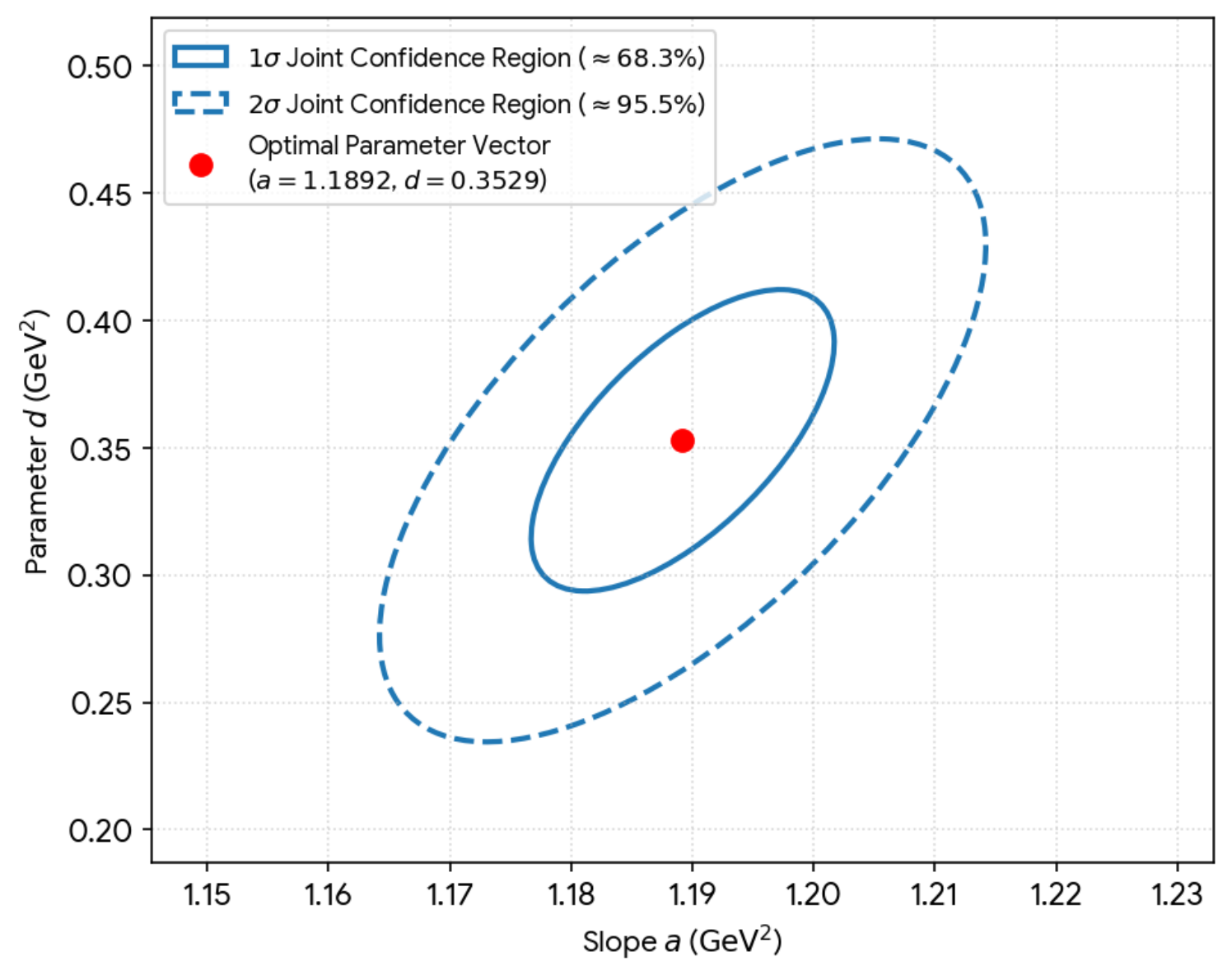}
    \caption{\small The joint confidence ellipses on the $(a, d)$ parameter plane for Model 3 evaluated on the global light meson database ($N=85$ states from Table~2). The red marker isolates the absolute nonlinear $\chi^2$ minimum. The solid and dashed contours plot the boundary positions of the $1\sigma$ ($\approx 68.3\%$) and $2\sigma$ ($\approx 95.5\%$) statistical confidence spaces, respectively.
    %The distinct vertical elongation underscores that the universal Regge slope $a$ is exceptionally well-constrained by the high-spin boundary conditions %compared to the short-range magnitude $d$, while the synchronized positive tilt reflects the stable linear parameter coupling ($r_{ad} = +0.647$) required to %stabilize the ground-state mass baseline.
    }
    \label{fig:error_ellipse}
\end{figure}

Since the radial/orbital slope $a$ is tightly constrained by the high-spin states ($J \le 6$), its absolute variance ($\sigma_a = 0.013\text{ GeV}^2$) is substantially smaller than the variance of the Dirac-Coulomb magnitude ($\sigma_d = 0.059\text{ GeV}^2$). As a result, the calculated $1\sigma$ standard error ellipse is highly elongated, with its major semi-axis $L_{\text{major}} \approx 0.061\text{ GeV}^2$ aligned almost vertically. The orientation angle of the ellipse relative to the $a$-axis is found to be $\phi \approx 88.5^\circ$, driven by the positive correlation coefficient $r_{ad} = +0.647$.

This highly directional tilt carries an important physical meaning: any marginal overestimation of the global string tension and its associated Regge slope $a$ during the fit forces a synchronized linear compensation in the short-range magnitude $d$ to maintain the mass scale of the lower trajectories. Despite this vertical elongation, the $1\sigma$ and $2\sigma$ confidence boundaries are completely closed and well-localized, proving that the absolute minimum is uniquely defined and free from flat-direction divergences (valleys) in the Hessian topology.

The correlation coefficient between the universal Regge slope $a$ and the Dirac-Coulomb magnitude $d$ is moderately bound ($r_{ad} = +0.647$). This proves that the parameters are statistically orthogonal and cleanly separated by the nonlinear fitting routine. The $d/(l+1)$ term does not merely absorb or mimic a global shift in the linear trajectory slope; instead, it models distinct physics that is concentrated strictly at low orbital angular momenta ($l=0,1$).

Conversely, we observe strong anti-correlations between the global constant intercept $c$ and the remaining parameters ($r_{ac} = -0.867$, $r_{dc} = -0.898$). This behavior is mathematically expected in multi-parameter regressions featuring an unconstrained scalar shift. Any systematic upward traction induced by larger values of the slope $a$ or the magnitude $d$ is automatically counterbalanced by the optimization algorithm via a lower intercept $c$. Despite this coupling, the Hessian matrix remains well-conditioned, ensuring that the computed parameter uncertainties are strictly stable.

The massive statistical significance of Model 3 on the full dataset is underscored by the Wald statistic ($Z \approx 6.0$), establishing the Dirac-Coulomb magnitude at an explicit $6\sigma$ confidence level ($d = 0.353 \pm 0.059 \text{ GeV}^2$). Furthermore, the model minimizes the required internal uncertainty ($\sigma_{\text{theor}} = 39.5 \text{ MeV}$) compared to the rigid linear regime ($\sigma_{\text{theor}} = 43.4 \text{ MeV}$), winning in the Bayesian Information Criterion ($\Delta\text{BIC} = \text{BIC}_{\text{Model 2}} - \text{BIC}_{\text{Model 3}} = 10.8$). On Jeffreys' scale for model selection (see Appendix~C), a $\Delta\text{BIC} > 10$ provides decisive evidence favoring the nonlinear correction.

\section{Technical Summary}

In this work, the modern Regge spectrum of light non-strange mesons has been systematically evaluated with the help of methods of mathematical statistics. The following parameterization of Regge-like trajectory was considered:
\begin{equation}
    \label{eq:regge_trajectory_new}
    m^2(n,l) = an + bl + \frac{d}{l+1} + c,
\end{equation}
where $n$ and $l$ are the radial and orbital angular momentum quantum numbers. We analyzed three specific configurations within the $(a,b,c,d)$ parameter space: Model~1 $(a,b,c,0)$, Model~2 $(a,a,c,0)$, and Model~3 $(a,a,c,d)$. Based on first-order error propagation, a nonlinear $\chi^2$ optimization scheme was implemented to self-consistently determine the intrinsic theoretical uncertainty $\sigma_{\text{theor}}$ under the exact constraint $\chi^2/\text{dof} = 1$. This approach avoids ad~hoc assumptions regarding model variances by utilizing the full Gaussian likelihood function. As a result, we demonstrate that a U-shaped minimum in the information criteria naturally emerges from the competition between $\chi^2$ minimization and the variance contribution of the covariance log-determinant.

Testing these models across different dataset profiles and error scales clarifies several aspects of the spectrum governed by nonperturbative QCD. Relaxing the slope degeneracy between radial and orbital excitations (Model~1) fails to provide a meaningful reduction in the required theoretical uncertainty $\sigma_{\text{theor}}$ for any of the samples. Statistical information criteria indicate that introducing independent slopes leads to parameter inflation without improving the fit quality, thereby confirming the stability of a universal Regge slope for light non-strange mesons.

For the benchmark subset of 27 well-established states, the pure linear trajectory exhibits noticeable tension, requiring a large intrinsic uncertainty of $91\text{~MeV}$. Introducing the Dirac-Coulomb term in Model~3 reduces this variance to $42\text{~MeV}$, with the extracted parameter $d = 0.410 \pm 0.097\text{~GeV}^2$ being significant at the $4.2\sigma$ level. A continuous scan over the error spectrum shows that the nonlinear trajectory consistently yields a lower information criterion minimum than the linear models.

When extended to the global dataset of 85 states, the trajectory parameters remain stable, and the significance of the Dirac-Coulomb term increases to the $6\sigma$ level ($d = 0.353 \pm 0.059\text{~GeV}^2$). On this expanded sample, Model~3 outperforms the linear models under both the Akaike and Bayesian Information Criteria, showing a clear statistical preference ($\Delta\text{BIC} = 10.8$) that remains robust against variations in the theoretical error scale.

Truncation tests performed by removing the $l=0$ states ($N=19$ for benchmarks and $N=73$ globally) verify the physical origin of the nonlinear correction. Excluding the $S$-waves regularizes the baseline linear trajectory and reduces the significance of the correction term to a marginal $3.0\sigma$ globally and an insignificant $1.7\sigma$ for the benchmark set. Concurrently, the Bayesian Information Criterion reverses its preference, indicating that the nonlinear term becomes redundant once short-range interactions are suppressed.

These results provide empirical support for the underlying QCD dynamics. The nonlinear correction to the Regge trajectories reflects short-range gluon exchange rather than overfitting. For states with higher angular momenta ($l \ge 1$), the centrifugal barrier suppresses short-range effects, leading to linear, stringlike behavior. Conversely, for the $S$-wave states, the short-range interaction induces a distinct distortion that must be accounted for phenomenologically.

\section{Conclusion}

The slopes of the radial and orbital Regge trajectories for light non-strange
mesons are identical within the experimental uncertainties of modern data.
Conversely, the alternative hypothesis of nonuniversal slopes lacks statistical
justification. The apparent nonuniversality observed in the data is driven
primarily by $S$-wave resonances, which induce a moderate deviation from
the strict linearity of both the orbital and radial trajectories. These observed
distortions can be uniformly parameterized by a Dirac-Coulomb-type nonlinear
correction. Physically, this implies that the Regge spectrum of light
non-strange mesons exhibits not only a Coulomb-like degeneracy but also a fine
splitting that arises from the first relativistic correction to the Coulomb problem.

The framework developed in this study provides an automated approach for
hadron classification, and the successful application of the considered nonlinear
parameterization motivates further extensions to heavy-light systems
and heavy quarkonia spectra, where short-range gluon interactions play a
dominant role in shaping the bound-state structure.

\appendix

\section*{Appendix A: Relativistic Correction to the Principal Quantum Number in the Coulomb Problem}

The specific nonlinear modification introduced in Model 3 is deeply rooted in relativistic quantum mechanics.
In a classical, non-relativistic Coulomb potential $V(r) = -Ze^2/r$, the discrete energy levels are strictly governed by the non-relativistic principal quantum number $n_{\text{Coul}}$,
\begin{equation}
    E \propto - \frac{1}{n^2_{\text{Coul}}},\qquad n_{\text{Coul}} = n_r + l + 1,
\end{equation}
where $n_r$ is the radial quantum number and $l$ is the orbital angular momentum quantum number.

However, for a relativistic spin-$1/2$ Dirac particle in a central Coulomb field, the exact energy spectrum is determined by the Sommerfeld-Dirac formula~\cite{dirac}. The structural behavior of this spectrum is controlled by an effective, relativistic principal quantum number $n_{\text{eff}}$,
\begin{equation}
\label{Neff}
    n_{\text{eff}} = n_r + \sqrt{\kappa^2 - (Z\alpha)^2},
\end{equation}
where $\alpha \approx 1/137$ is the fine-structure constant, and the relativistic angular quantum number $\kappa$ is related to the total angular momentum $j$ via $|\kappa| = j + 1/2$. For states with the maximum total angular momentum for a given $l$, namely $j = l + 1/2$, one has $|\kappa| = l + 1$. Although the exact energy levels for $j = l - 1/2$ coincide with those of $j = l - 1/2$ from the adjacent shell due to relativistic degeneracy, the structural form of the correction is conventionally illustrated using the $j = l + 1/2$ branch. Substituting $|\kappa| = l + 1$ into~\eqref{Neff} and expanding in powers of $\alpha$ yield
\begin{equation}
\label{Neff2}
    n_{\text{eff}} = n_r + (l+1)\sqrt{1 - \frac{(Z\alpha)^2}{(l+1)^2}} \approx n_r + l + 1 - \frac{(Z\alpha)^2}{2(l+1)}.
\end{equation}
The leading relativistic correction --- the last term in~\eqref{Neff2} --- decreases the effective quantum number, thereby making the discrete energy states more deeply bound. This effect is most pronounced for the $l=0$ states and leads to the fine-structure splitting of energy levels that possess identical $n_{\text{Coul}}$ but different $l$.

In the spectrum of light non-strange mesons, an approximate degeneracy formally reminiscent of the Coulomb-type behavior emerges in the form $m^2 \propto n_{\text{Coul}}$, where $n_{\text{Coul}} = n_r + l + 1$ (with $n_r$ reparameterized as $n$ in the standard regge notation). It is therefore natural to assume that the leading correction to this phenomenological law inherits the functional form of~\eqref{Neff2}. This assumption leads to the ansatz of Model~3:
\begin{equation}
\label{relcor}
    m^2(n,l) = a(n + l) + c + \frac{d}{l+1},
\end{equation}
For $d > 0$, the nonlinear correction  gives an additional contribution to the resonance mass. This effect is especially pronounced for the $S$-wave ($l=0$) states.

The nonlinear correction in~\eqref{relcor} can be interpreted as a relativistic modification to the orbital trajectory slope, arising naturally in relativized quark models with $d \propto \alpha_s^2$~\cite{Jia:2017age}, which is in agreement with~\eqref{Neff2}.

\section*{Appendix B: Analytical Expressions for the Hessian Matrix Elements}

The Hessian matrix $H$ allows to rigorously quantify the statistical uncertainties and mutual dependencies of the extracted model parameters.
We will consider the example of Model 3, the corresponding set of parameters is denoted as $\boldsymbol{\theta} = \{a, d, c\}$.
The Hessian matrix here is a square $3 \times 3$ symmetric matrix comprising the second-order partial derivatives of the objective $\chi^2$ function evaluated at the local minimum:
\begin{equation}
\label{hes}
    H_{jk} = \frac{\partial^2 \chi^2}{\partial \theta_j \partial \theta_k} =
    \begin{pmatrix}
        \frac{\partial^2 \chi^2}{\partial a^2} & \frac{\partial^2 \chi^2}{\partial a \partial d} & \frac{\partial^2 \chi^2}{\partial a \partial c} \\
        \frac{\partial^2 \chi^2}{\partial d \partial a} & \frac{\partial^2 \chi^2}{\partial d^2} & \frac{\partial^2 \chi^2}{\partial d \partial c} \\
        \frac{\partial^2 \chi^2}{\partial c \partial a} & \frac{\partial^2 \chi^2}{\partial c \partial d} & \frac{\partial^2 \chi^2}{\partial c^2}
    \end{pmatrix}.
\end{equation}

Geometrically, the Hessian matrix characterizes the multidimensional curvature
of the parameter space in the vicinity of the optimal solution. In statistical
inference and regression analysis, the Hessian serves three critical purposes
that determine the validity of the phenomenological fit. First, it provides
an exact framework for parameter uncertainty quantification, as the local
curvature of the $\ln \mathcal{L}$ (or $\chi^2$) surface directly determines
the statistical covariance of the optimized parameters. A steep minimum,
characterized by high curvature and large second-order partial derivatives,
implies that even a small deviation in the parameter vector $\boldsymbol{\theta}$
yields a substantial reduction in the log-likelihood, indicating tight parameter
constraints. Conversely, a shallow minimum signifies large statistical uncertainties.
Mathematically, the parameter covariance matrix $V$ is obtained via direct matrix inversion of the Hessian,
\begin{equation}
    V = H^{-1},
\end{equation}
from which the standard errors for each individual parameter are subsequently extracted as the square roots of the diagonal entries:
\begin{equation}
    \sigma_a = \sqrt{V_{aa}}, \quad \sigma_d = \sqrt{V_{dd}}, \quad \sigma_c = \sqrt{V_{cc}}.
\end{equation}

Second, the Hessian matrix allows for a rigorous assessment of parameter
correlations and mixing. The normalized off-diagonal elements of the
covariance matrix $V$ define the linear correlation coefficients,
$r_{jk} = V_{jk}/\sqrt{V_{jj} V_{kk}}$. For Model~3, extracting these
coefficients is essential to verify whether the nonlinear trajectory
modification is genuinely independent or merely mimics a global scale
transformation; a moderate correlation confirms that the parameters describe
distinct physical regimes.

Finally, the Hessian framework is indispensable for verifying local
convergence. To guarantee that the optimization routine has successfully
converged to a genuine local minimum rather than a saddle point or a numerical
artifact, the Hessian matrix must be strictly positive-definite. This topological
condition ensures that all eigenvalues of $H$ are strictly positive,
confirming that any perturbation away from the optimized vector
$\boldsymbol{\theta}$ monotonically decreases the log-likelihood function.

Consider the Hessian matrix~\eqref{hes} in our specific case. The functional form of the $\chi^2$ from Eq.~\eqref{eq:chi2} can be rewritten for convenience as:
\begin{equation}
    \chi^2 = \sum_{i=1}^{N} \frac{\Delta_i^2}{w_i},
\end{equation}
where the residuals $\Delta_i$ and the fixed statistical weights $w_i$ are given by
\begin{equation}
    \Delta_i = m_{i,\text{exp}}^2 - a(n_i + l_i) - \frac{d}{l_i+1} - c,
\end{equation}
\begin{equation}
    w_i = 4m_{i,\text{exp}}^2\left(\sigma_{i,\text{pdg}}^2 + \sigma_{\text{theor}}^2\right).
\end{equation}

The first-order partial derivatives of $\chi^2$ with respect to each parameter are
\begin{equation}
    \frac{\partial \chi^2}{\partial a} = -2 \sum_{i=1}^{N} \frac{\Delta_i (n_i + l_i)}{w_i},
\end{equation}
\begin{equation}
    \frac{\partial \chi^2}{\partial d} = -2 \sum_{i=1}^{N} \frac{\Delta_i}{w_i (l_i + 1)},
\end{equation}
\begin{equation}
    \frac{\partial \chi^2}{\partial c} = -2 \sum_{i=1}^{N} \frac{\Delta_i}{w_i}.
\end{equation}

Differentiating a second time yields the elements of the symmetric $3 \times 3$ Hessian matrix~\eqref{hes}.
Its diagonal elements define the pure curvature of the parameter space and are strictly positive,
\begin{equation}
    H_{aa} = \frac{\partial^2 \chi^2}{\partial a^2} = 2 \sum_{i=1}^{N} \frac{(n_i + l_i)^2}{w_i},
\end{equation}
\begin{equation}
    H_{dd} = \frac{\partial^2 \chi^2}{\partial d^2} = 2 \sum_{i=1}^{N} \frac{1}{w_i (l_i + 1)^2},
\end{equation}
\begin{equation}
    H_{cc} = \frac{\partial^2 \chi^2}{\partial c^2} = 2 \sum_{i=1}^{N} \frac{1}{w_i}.
\end{equation}

The off-diagonal elements govern the linear correlations and mixings between the phenomenological trajectory components,
\begin{equation}
    H_{ad} = \frac{\partial^2 \chi^2}{\partial a \partial d} = 2 \sum_{i=1}^{N} \frac{n_i + l_i}{w_i (l_i + 1)},
\end{equation}
\begin{equation}
    H_{ac} = \frac{\partial^2 \chi^2}{\partial a \partial c} = 2 \sum_{i=1}^{N} \frac{n_i + l_i}{w_i},
\end{equation}
\begin{equation}
    H_{dc} = \frac{\partial^2 \chi^2}{\partial d \partial c} = 2 \sum_{i=1}^{N} \frac{1}{w_i (l_i + 1)}.
\end{equation}

The covariance matrix $V$ is subsequently computed via direct matrix inversion $V = H^{-1}$, from which the parameter errors are extracted as $\sigma_{\theta_j} = \sqrt{V_{jj}}$, and the correlation coefficients are defined as $r_{jk} = V_{jk}/\sqrt{V_{jj} V_{kk}}$.

\section*{Appendix C: Information Criteria}

When comparing models where the theoretical uncertainty $\sigma_{\text{theor}}$
is treated as a free or dynamically adjusted parameter, utilizing the
$\chi^2$ minimum value for model selection is statistically insufficient.
Instead, model comparison must be based on the maximization of the full
Gaussian log-likelihood function, $\ln \mathcal{L}$. Under the assumption of
independent, normally distributed experimental uncertainties, the total
likelihood $\mathcal{L}$ for observing the dataset $\mathbf{m}^2_{\text{exp}}$
given the parameters $\boldsymbol{\theta}$ and $\sigma_{\text{theor}}$ is
expressed as
\begin{equation}
    \mathcal{L} = \prod_{i=1}^{N} \frac{1}{\sqrt{2\pi w_i}} \exp\left( -\frac{\Delta_i^2}{2w_i} \right),
\end{equation}
where $w_i = 4m_{i,\text{exp}}^2\left(\sigma_{i,\text{pdg}}^2 + \sigma_{\text{theor}}^2\right)$ represents the variance of the squared mass as defined in the previous section. Taking the natural logarithm of both sides yields the log-likelihood function,
\begin{equation}
    \ln \mathcal{L} = -\frac{N}{2}\ln(2\pi) - \frac{1}{2} \sum_{i=1}^{N} \ln(w_i) - \frac{1}{2}\chi^2,
\end{equation}
where the argument of the logarithm is made dimensionless by an implicit normalization constant.

Omitting the constant additive term $-\frac{N}{2}\ln(2\pi)$, which cancels out identically during model differentiation, we define the effective negative log-likelihood score as
\begin{equation}
    -2\ln \mathcal{L}_{\text{eff}} = \chi^2 + \sum_{i=1}^{N} \ln\left[4m_{i,\text{exp}}^2\left(\sigma_{i,\text{pdg}}^2 + \sigma_{\text{theor}}^2\right)\right].
\end{equation}

The Akaike Information Criterion (AIC), which is derived from minimizing the Kullback-Leibler (KL) divergence between the true data distribution and the model, introduces an asymptotic penalty for the number of free parameters $k$:
\begin{equation}
    \text{AIC} = -2\ln \mathcal{L}_{\text{eff}} + 2k = \chi^2 + \sum_{i=1}^{N} \ln\left[4m_{i,\text{exp}}^2\left(\sigma_{i,\text{pdg}}^2 + \sigma_{\text{theor}}^2\right)\right] + 2k.
\end{equation}

The Bayesian Information Criterion (BIC), derived from a limiting approximation of the Bayesian posterior probability under a uniform prior, imposes a stricter penalty scaled by the logarithm of the total sample size $N$:
\begin{equation}
    \text{BIC} = -2\ln \mathcal{L}_{\text{eff}} + k\ln(N) = \chi^2 + \sum_{i=1}^{N} \ln\left[4m_{i,\text{exp}}^2\left(\sigma_{i,\text{pdg}}^2 + \sigma_{\text{theor}}^2\right)\right] + k\ln(N).
\end{equation}

Within the framework of model selection and information theory, both the AIC and BIC serve to evaluate competing models by balancing goodness of fit against parameter inflation. Despite their structural similarities, they derive from fundamentally distinct theoretical principles and address different asymptotic goals regarding the underlying data-generating distribution.
The AIC is dynamically rooted in information theory and represents an asymptotically unbiased estimator of the expected KL divergence between the true physical data and the approximating model. This is a practical way to estimate the KL divergence in conditions where we do not know the true distribution of the data. It is widely used in machine learning. This information criterion assumes that the true data-generating mechanism is infinitely complex, thereby aiming to select the model that minimizes information loss for future predictive data.
Conversely, the BIC operates under the assumption that the true, exact model is contained within the candidate set. Consequently, the BIC introduces a heavier, sample-size-dependent penalty on model complexity for any dataset where $\ln(N) > 2$.
The choice between these criteria represents a fundamental trade-off in statistical inference. The AIC is structurally optimized for predictive performance and tends to favor more complex parameter spaces to avoid underfitting. In contrast, the BIC exhibits mathematical consistency, asymptotically identifying the true lower-dimensional model as $N \rightarrow \infty$, making it the preferred choice for physical applications requiring strict parameter parsimony and model interpretability.

In our specific workflow, because $\sigma_{\text{theor}}$ is adjusted to force $\chi^2 = N - k$, the explicit values reduce to
\begin{multline}
    \text{AIC} = (N - k) + \sum_{i=1}^{N} \ln\left[4m_{i,\text{exp}}^2\left(\sigma_{i,\text{pdg}}^2 + \sigma_{\text{theor}}^2\right)\right] + 2k = \\
    N + k + \sum_{i=1}^{N} \ln\left[4m_{i,\text{exp}}^2\left(\sigma_{i,\text{pdg}}^2 + \sigma_{\text{theor}}^2\right)\right],
\end{multline}
\begin{equation}
    \text{BIC} = N - k + \sum_{i=1}^{N} \ln\left[4m_{i,\text{exp}}^2\left(\sigma_{i,\text{pdg}}^2 + \sigma_{\text{theor}}^2\right)\right] + k\ln(N).
\end{equation}
As a consequence, the model that yields a lower value of $\sigma_{\text{theor}}$ directly gains a massive numerical advantage inside the summation term $\sum \ln(\dots)$. For instance, \(\sigma _{\text{theor}}\) of Model~3 is noticeably smaller than that of competing models, and the corresponding logarithmic term \(\sum \ln(\dots)\) takes on a significantly lower (more negative) value. This gain more than outweighs the penalty for introducing the additional parameter \(d\).

Finally, if the BIC of two conditional models 1 and 2 is known, the difference $\Delta\text{BIC} = \text{BIC}_{\text{model 2}} - \text{BIC}_{\text{model 1}}$ is often exploited as a quantitative criterion for model selection with the help of the so-called Jeffreys' scale. Unlike the standard \(p\)-value, the Jeffreys' scale takes into account the a priori uncertainty of the parameters of physical models and penalizes for excessive complexity (overfitting). It provides a rigorous, prior-weighted alternative to frequentist testing. Technically, the method is based on the known asymptotic approximation for large samples: $\Delta\text{BIC} \approx 2\ln(B_{12})$, where $B_{12}$ is the Bayes factor. According to Jeffreys' scale, $0 < \ln(B_{12}) < 1$ indicates inconclusive evidence, $1 < \ln(B_{12}) < 2.5$ weak evidence, $2.5 < \ln(B_{12}) < 5$ strong evidence, and $\ln(B_{12}) > 5$ decisive evidence in favor of Model 1. The mapping to $\Delta\text{BIC}$ is easily accomplished by multiplying by 2. For example, the strong evidence is achieved at $5 \lesssim \Delta\text{BIC} \lesssim 10$.

\end{document}